\documentclass[reqno,12pt,draft]{amsart} 

\usepackage{amsmath}
\usepackage{amsfonts}
\usepackage{enumerate}
\usepackage{fancybox}

\newcommand{\p}{\partial}

\theoremstyle{plain}
\newtheorem{thm}{Theorem}[section]{\bf}{\it}
\newtheorem{prop}[thm]{Proposition}{\bf}{\it}
\newtheorem{lemma}[thm]{Lemma}{\bf}{\it}
\newtheorem{lem}[thm]{Lemma}{\bf}{\it}
{\bf}{\it}
\theoremstyle{definition}
\newtheorem*{acknowledgement}{Acknowledgement} 
\newtheorem{remark}[thm]{Remark}{\it}{\rm}
{\bf}{\rm}
{\rm}{\rm}

{\bf}{\rm}
\newcommand{\supp}{{\operatorname{supp}}}

\newcommand{\Vol}{{\operatorname{Vol}}}

\newenvironment{pf*}[1]{\par\medskip\noindent\textit{#1}\,:}{\hspace*{\fill}\qed\medskip\par\noindent}  

\numberwithin{equation}{section}

\newcommand{\R}{{\mathbb R}}
\newcommand{\N}{{\mathbb N}}
\newcommand{\C}{{\mathbb C}}

\title[Analytic structure of Coulombic wave functions]{Analytic
  structure of many-body Coulombic wave functions}  

\thanks{\copyright\ 2008 by the
       authors. This article may be reproduced in its entirety for
       non-commercial purposes.}

\author[S. Fournais, M. and T. Hoffmann-Ostenhof, and T. \O. S\o rensen]
{S. Fournais \and M. Hoffmann-Ostenhof \and T. Hoffmann-Ostenhof \and
T. \O stergaard S\o rensen}

\address[S. Fournais]{Department of Mathematical Sciences, University 
  of Aarhus, Ny Munkegade, Building
  1530, DK-8000 \AA rhus C, Denmark.}
\email{fournais@imf.au.dk}           
\address[S. Fournais on leave from]{CNRS and Laboratoire de
  Math\'{e}matiques d'Orsay, 
Univ Paris-Sud, Orsay CEDEX, F-91405, France.}

\address[M. Hoffmann-Ostenhof]
        {Fakult\"at f\"ur Mathematik,
         Universit\"at Wien,         
         Nordbergstra\ss e 15,
         A-1090 Vienna,
         Austria.} 
\email{maria.hoffmann-ostenhof@univie.ac.at}

\address[T. Hoffmann-Ostenhof]{Institut f\"ur Theoretische
Chemie, W\"ahringer\-strasse 17,
           Universit\"at Wien,
           A-1090 Vienna,
           Austria.}

\address[T. Hoffmann-Ostenhof, 2nd address]{
        The Erwin Schr\"{o}dinger International Institute for 
        Mathematical Physics,
              Boltzmanngasse 9,
              A-1090 Vienna, Austria.}
\email{thoffman@esi.ac.at}

\address[T. \O stergaard S\o rensen]
{Department of Mathematical Sciences,
           Aalborg University,
           Fredrik Bajers Vej 7G,
           DK-9220 Aalborg East, Denmark.}
\email{sorensen@math.aau.dk}

\date{\today}

\begin{document}

\thispagestyle{empty}

\begin{abstract}
We investigate the analytic structure of solutions of non-relativistic 
Schr\"odinger equations describing Coulombic many-par\-ticle
systems. We prove the following: Let $\psi({\bf x})$ with ${\bf
  x}=(x_1,\dots, x_N)\in 
\mathbb R^{3N}$ denote an $N$-electron wavefunction of such a system
with one nucleus fixed at the origin. Then in a neighbourhood of a
coalescence point, for which $x_1=0$ and the other electron
coordinates  do not coincide, and differ from 0, $\psi$  can be
represented locally as $\psi({\bf x})=\psi^{(1)}({\bf x})+|x_1|\psi^{(2)}({\bf
  x})$ with
$\psi^{(1)},\psi^{(2)}$ real analytic. A similar representation holds
near 
two-electron coalescence points. 
The Kustaanheimo-Stiefel transform and analytic hypoellipticity
play an essential role in the
proof. 
\end{abstract} 

\maketitle

\section{Introduction and results}
\subsection{Introduction}
In quantum chemistry and atomic and molecular physics, the regularity
properties of the Coulombic  wavefunctions \(\psi\), and of their
corresponding one-electron densities \(\rho\), are of great
importance. These regularity properties determine  the convergence
properties of various (numerical) approximation schemes (see
\cite{BrisLions, Flad1, Flad2, y1,y2,y3} for some recent works). 
They  are also of intrinsic mathematical interest. 

The pioneering work is due to Kato \cite{kato57}, who proved that
$\psi$ is Lipschitz continuous, i.e., $\psi\in C^{0,1}$, near
two-particle coalescence points.

In a series of recent papers the present authors have studied these
properties in  detail. In \cite{CMP2} we deduced an optimal
representation of $\psi$ of the form $\psi=\mathcal F\Phi$ with an
explicit $\mathcal F\in C^{0,1}$, such that $\Phi\in C^{1,1}$,
characterizing the singularities of $\psi$ up to second derivatives;
see \cite[Theorem 1.1]{CMP2} for a precise statement. In particular,
$\mathcal F$ contains logarithmic terms which stem from the
singularities of the potential at three-particle coalescence
points. This characterization has been applied in \cite{AHP2} and
\cite{3der} in the study of the electron density $\rho$ and (in the
atomic case) its spherical average $\widetilde\rho$ close to the
nuclei. Real analyticity of $\rho$ away from the nuclei was proved in
\cite{Ark}; see also \cite{CMP1}, \cite{Taxco}.  

In this paper we derive a different representation of \(\psi\) which
completely settles its analytic structure in the neighbourhood of
two-par\-ticle coalescence points.
The Kustaanheimo-Stiefel transform (KS-transform for short) and
analytic hypoanalyticity of a certain degenerate elliptic operator are
crucial for the proof.  

We start with the one-particle case.

\begin{thm}\label{Hill}
Let $\Omega\subset \mathbb R^3$ be a neighbourhood of the origin and
assume that $W^{(1)},\:W^{(2)},\: F^{(1)}$, and $F^{(2)}$ are real
analytic functions in $\Omega$. Let  
\begin{equation}\label{Hone}
  H=-\Delta+\frac{W^{(1)}}{|x|}+W^{(2)}\,,
\end{equation}
and assume that  \(\varphi\in W^{1,2}(\Omega)\) satisfies 
\begin{equation}\label{Hpsi}
  H\varphi=\frac{F^{(1)}}{|x|}+F^{(2)}
\end{equation}
in $\Omega$ in the distributional sense.

Then there exists a neighbourhood \(\widetilde\Omega\subset\Omega\) of
the origin, and real analytic functions $\varphi^{(1)},
\varphi^{(2)}:\widetilde\Omega\to\C\) such that
\begin{equation}\label{psian}
  \varphi(x)=\varphi^{(1)}(x)+|x|\varphi^{(2)}(x)\,, \ x\in\widetilde\Omega\,.
\end{equation}
\end{thm}
\begin{remark}\label{remark1}
Theorem \ref{Hill} is a generalization of an almost 25 years old
result by Hill  \cite{Hill}. The present investigations were partly
motivated by this work. Hill considered solutions to 
\begin{equation}\label{Hill1}
  \big(-\Delta-\frac{Z}{|x|}+V^{(1)}(x)+|x|V^{(2)}(x)\big)\varphi=0\,, 
\end{equation}
with $V^{(1)}$ and $V^{(2)}$ real analytic near the origin, and proved
that $\varphi$ satisfies \eqref{psian}. The statement \eqref{psian} can
easily be seen to hold for Hydrogenic eigenfunctions.  These have the
form $e^{-\beta |x|}P(x)$ for some $\beta>0$, where $P(x)$ can be
written as linear combinations  of polynomials in $|x|$ times
homogeneous harmonic polynomials. In particular, Hill's result implies
that $\varphi$ satisfies \eqref{psian} near the origin for a one-electron
molecule with fixed nuclei, one of them at the origin.  
\end{remark}

\begin{remark}\label{remark2}
Hill's proof is rather involved. Our proof is quite different, also
not easy, but has the advantage that it can be generalized to treat
the Coulombic many-particle case; see Theorem \ref{thm:main} and its
proof below,
and also Remark~\ref{rem:1}.

The proof of Theorem~\ref{Hill} uses the KS-transform (see
Section~\ref{sec:Hill} for the definition). This transform  was
introduced in the 1960's \cite{KS} to regularize the Kepler problem in
classical mechanics (see also \cite{Kustaanheimo,stiefel, Knauf}) and
has found applications in problems related to the Coulomb potential in
classical mechanics and quantum mechanics, see
\cite{JeckoKnauf,GerardKnauf, HelfferEtAl, HelfferSiedentop1,
  HelfferSiedentop2, Jost}. The KS-transform is a homogeneous
extension of the Hopf map (also called the Hopf fibration), the first
example of a map from $\mathbb S^3$ to $\mathbb S^2$ which is not
null-homotopic, discovered in the 1930's \cite{Hopf}. For more on the
literature on the KS-transform, see \cite{Knauf, HelfferSiedentop1}.
\end{remark}

We move to the \(N\)-particle problem. For the sake of simplicity we
consider the atomic case and  mention extensions in the remarks. 
Let $H$ be the non-relativistic Schr\"odinger operator of an
$N$-electron atom with nuclear charge $Z>0$ in the fixed nucleus  
approximation,
\begin{equation}\label{H}
  H=\sum_{j=1}^N\Big(-\Delta_j-\frac{Z}{|x_j|}\Big)
    +\sum_{1\le i<j\le N}\frac{1}{|x_i-x_j|}\,=:-\Delta + V\,.
\end{equation}
Here the $x_j=(x_{j,1},x_{j,2},x_{j,3})\in \mathbb R^3$, $j=1,\dots,
N$, denote the positions of the electrons, and the $\Delta_j$ are the
associated Laplacians so that $\Delta=\sum_{j=1}^N\Delta_j$ is the
$3N$-dimensional Laplacian. Let ${\bf x}=(x_1,x_2,\dots, x_N)\in
\mathbb R^{3N}$ and let $\nabla=(\nabla_1,\dots, \nabla_N)$ 
denote the $3N$-dimensional gradient operator. 
The operator $H$ is bounded from \(W^{2,2}(\R^{3N})\) to
\(L^2(\R^{3N})\), and defines a bounded quadratic form on
\(W^{1,2}(\R^{3N})\) \cite{katobook}.
%
%
We investigate local 
solutions $\psi$ of  
\begin{align}\label{eigen}
  H\psi=E\psi\ , \quad E\in \R\,,
\end{align} 
in a neighbourhood of  two-particle coalescence points.

More precisely, let \(\Sigma\) denote the set of coalescence points, 
\begin{align}\label{eq:Sigma}
  \Sigma:=\Big\{{\bf x}=(x_1,\ldots,x_N)\in\R^{3N}\,\Big|\,
          \prod_{j=1}^{N}|x_j|
          \!\!\prod_{1\le i<j\le N}|x_i-x_j|=0\Big\}\,.
\end{align}

If, for some \(\Omega\subset\R^{3N}\), \(\psi\) is a distributional  
solution to \eqref{eigen} in \(\Omega\), then \cite[Section 7.5,
pp.\ 177--180]{Hormander} \(\psi\) is real
analytic away from \(\Sigma\), that is, \(\psi\in
C^{\omega}(\Omega\setminus\Sigma)\).

Let, for \(k,\ell\in\{1,\ldots,N\}, k\neq \ell\), 
\begin{align}
  \label{eq:Sigma-k}
  \Sigma_k':&=\Big\{{\bf x}\in\R^{3N}\,\Big|\,
           \prod_{j=1,j\neq k}^{N}|x_j|\prod_{1\le i<j\le
             N}|x_i-x_j|=0\Big\}\,,\\
  \label{eq:Sigma-k-l}
  \Sigma_{k,\ell}':&=\Big\{{\bf x}\in\R^{3N}\,\Big|\,
           \,\prod_{j=1}^{N}|x_j|\prod_{1\le i<j\le
             N, \{i,j\}\neq\{k,\ell\}}
          \!\!\!\!\!\!\!|x_i-x_j|=0\Big\}\,.
\end{align}
Then we denote
\begin{align}
  \label{eq:sigmas}
  \Sigma_k:=\Sigma\setminus\Sigma_k'\ , \quad
  \Sigma_{k,\ell}:=\Sigma\setminus\Sigma_{k,\ell}'
\end{align}
the two kinds of `two-particle coalescence
points'. 

The main result of this paper is the following.
 \begin{thm}\label{thm:main}
Let \(H\) be the non-relativistic Hamiltonian of an atom, given by
\eqref{H}, let \(\Omega\subset\R^{3N}\) be an open set, and
assume that $\psi\in W^{1,2}(\Omega)$ satisfies, for some
\(E\in\R\), 
\begin{align}
  \label{eq:eigen}
  H\psi=E\psi\ \ \text{ in }\ \Omega
\end{align}
in the distributional sense. 
Let the sets \(\Sigma_k\) and
\( \Sigma_{k,\ell}\) be given by \eqref{eq:sigmas}.

Then, for all \(k\in\{1,\ldots,N\}\), there exists a neighbourhood
\(\Omega_k\subset\Omega\) of \(\Omega\cap\Sigma_k\), and
real analytic functions
\(\psi_k^{(1)},\psi_{k}^{(2)}:\Omega_k\to\C\) such that 
\begin{align}
  \label{eq:resultPSIa}
  &\psi({\bf x})=\psi_{k}^{(1)}({\bf x})+|x_k|\psi_{k}^{(2)}({\bf
    x})\,,\ {\bf x}\in\Omega_k\,,
\end{align}
and for all \(k,\ell\in\{1,\ldots,N\}\), \(k\neq\ell\),  there exists
a neighbourhood \(\Omega_{k,\ell}\subset\Omega\) of
\(\Omega\cap\Sigma_{k,\ell}\), and real analytic functions
\(\psi_{k,\ell}^{(1)},\psi_{k,\ell}^{(2)}:\Omega_{k,\ell}\to\C\) such
that 
\begin{align}
  \label{eq:resultPSIb}
  &\psi({\bf x})=\psi_{k,\ell}^{(1)}({\bf
    x})+|x_k-x_\ell|\psi_{k,\ell}^{(2)}({\bf x})\,,\ {\bf x}\in\Omega_{k,\ell}\,. 
\end{align}
\end{thm}
\begin{remark}\label{rem:0}
The proof of Theorem \ref{thm:main} again uses the KS-transform. Due
to the presence of the other electron coordinates we are confronted
with additional problems. We have to deal with degenerate elliptic
PDE's where the corresponding operators (of Gru{\v{s}}in-type) are
analytic  hypoelliptic, see \cite{grusin2}.
\end{remark}
\begin{remark}\label{rem:1}
  Theorem~\ref{thm:main} extends in the obvious way to
  electronic eigenfunctions of Hamiltonians of $N$-electron molecules
  with  \(K\) nuclei fixed at positions $(R_1,\ldots, R_K) \in \mathbb
  R^3$, given by  
\begin{equation}\label{moleculfixed}
  H=\sum_{j=1}^N\Big(-\Delta_j
  -\sum_{\ell=1}^K\frac{Z_\ell}{|x_j-R_\ell|}\Big)
  +\sum_{1\le i<j\le N}\frac{1}{|x_i-x_j|}\,.
\end{equation}

Furthermore we can replace  in \eqref{moleculfixed}, as in
Theorem~\ref{Hill}, the potential terms by more general terms, and
allow for inhomogeneities.   

For instance, the result holds for general Coulombic 
many-particle systems described by 
\begin{equation}\label{Hv}
 H=\sum_{j=1}^n{}-\frac{\Delta_j}{2m_j}+\sum_{1\le i<j\le
   n}v_{ij}(x_i-x_j)\,, 
\end{equation}
where the $m_j>0$ denote the masses of the particles, and 
$v_{ij}=v_{i,j}^{(1)}|x_i-x_j|^{-1}+v^{(2)}_{i,j}$ with 
$v^{(k)}_{ij},\:k=1,2$, real analytic. 
\end{remark}

\begin{remark}\label{extensions}
In separate work we will present additional regularity results (not
primarily for Coulomb problems) obtained partly using the techniques
developed in the present paper. 
\end{remark}
\section{Proofs of the main theorems}
\label{sec:Hill}
As mentioned in the introduction our proofs are based on the
Ku\-staan\-heimo-Stiefel (KS) transform. We will 'lift' the
differential equations to new coordinates using that transform. The
solutions to the new equations will be real analytic functions. By
projecting to the original coordinates we get the structure results
Theorem~\ref{Hill} and Theorem~\ref{thm:main}. 

In the present section we will introduce the KS-transform and show how
it allows to obtain Theorems~\ref{Hill} and \ref{thm:main}. The
more technical verifications of the properties of the KS-transform and its
composition with real analytic functions needed for these proofs are left
to Sections~\ref{sec:KS} and \ref{sec:AnAndKS}. 

Define the KS-transform \(K:\R^4\to\R^3\) by 
\begin{align}
  \label{eq:KSbis}
  K(y)=\left(\begin{matrix}
   y_1^2-y_2^2-y_3^2+y_4^2\\
   2(y_1y_2-y_3y_4)\\
   2(y_1y_3+y_2y_4)
   \end{matrix}\right)\,.
\end{align}
It is a simple computation to verify that
\begin{align}
  \label{eq:norms}
  |K(y)|:=\|K(y)\|_{\R^3} = \|y\|_{\R^4}^2=:|y|^2\,  \text{ for all
  }y\in\R^4\,. 
\end{align}
Let \(f:\R^3\to\C\) be any
\(C^2\)-function, and define, with \(K\) as above,
\begin{align}\label{eq:def-compo}
  f_K:\R^4\to\C\ , \qquad f_K(y):=f(K(y))\,.
\end{align}
Then for all \(y\in\R^4\setminus\{0\},\) (see Lemma~\ref{lem:2}),
\begin{align} \label{eq:Laplace-formula}
  (\Delta f)(K(y))&=\frac{1}{4|y|^2}\,\Delta f_K(y)\,.
\end{align}
\subsection{Proof of Theorem~\ref{Hill}}
Assume \(\varphi\in W^{1,2}(\Omega)\) satisfies (see
\eqref{Hone}--\eqref{Hpsi}) 
\begin{align}
  \label{eq:one-partBIS}
  \big({}-\Delta+\frac{W^{(1)}}{|x|}+W^{(2)}\big)\varphi
  =\frac{F^{(1)}}{|x|}+F^{(2)}\,, 
\end{align}
with \(W^{(1)}, W^{(2)}, F^{(1)}, F^{(2)}\) real analytic in
\(\Omega\subset\R^3\). 


Assume without loss that \(\Omega=B_3(0,r)\) for some \(r>0\). 
(Here, and in the sequel, \(B_n(x_0,r)=\{x\in\R^n\,|\, |x-x_0|<r\}\).)
Since
\(\varphi\in L^2(\Omega)\), 
Remark~\ref{rem:ext-K} in Section~\ref{sec:KS} below implies that  
\(\varphi_K\) is well-defined, as an element of
\(L^2(K^{-1}(\Omega),\tfrac{4}{\pi}|y|^2dy)\). 
We will show that \(\varphi_K\) satisfies (in the distributional sense)
\begin{align}
  \label{eq:new-psi-y}
  \big({}-\Delta_y+4(W^{(1)}_K+|y|^2W^{(2)}_K)\big)\varphi_K
  =4(F^{(1)}_K+|y|^2F^{(2)}_K)\,
\end{align}
in \(K^{-1}(\Omega)=B_4(0,\sqrt{r})\), with \(W^{(i)}_K, F^{(i)}_K\), \(i=1,2\),
defined as in \eqref{eq:def-compo}. Since \(W^{(i)}, F^{(i)}\),
\(i=1,2\), are real analytic in \(B_3(0,r)\) by assumption, and
\(K:\R^4\to\R^3\) (see \eqref{eq:KSbis}) and \(y\mapsto|y|^2\) are
real analytic, the coefficients in the elliptic equation in
\eqref{eq:new-psi-y} are real analytic in \(B_4(0,\sqrt{r})\).
It follows from elliptic regularity for equations with real analytic
 coefficients \cite[Section 7.5,
 pp.\ 177--180]{Hormander}
that \(\varphi_K:B_4(0,\sqrt{r})\to\C\) is real analytic. The statement of 
Theorem~\ref{Hill} then follows from 
Proposition~\ref{prop:one particle} in Section~\ref{sec:AnAndKS}
below. 

It therefore remains to prove that \(\varphi_K\) satisfies
\eqref{eq:new-psi-y}. 

By elliptic regularity,  $\varphi \in
W^{2,2}(\Omega')$ for all \(\Omega'=B_3(0,r')\), $r' < r$. (To see
this, use Hardy's inequality \cite[Lemma p.\ 169]{RS2} and that \(\varphi\in
W^{1,2}(\Omega)\) to conclude that \(\Delta\varphi=G\) with \(G\in
L^{2}(\Omega')\). Then use \cite[Theorem 8.8]{GT}). 

It follows 
that both \((\Delta\varphi)_K\)
 and \((|\,\cdot\,|^{-1}\varphi)_K\) are well-defined, as elements of
\(L^2(K^{-1}(\Omega'),\tfrac{4}{\pi}|y|^2dy)\) (see Remark~\ref{rem:ext-K} in
 Section~\ref{sec:KS} 
 below; see also \eqref{eq:L-2-abs}). This and
\eqref{eq:one-partBIS} imply that, as functions in
\(L^2(K^{-1}(\Omega'))=L^2(B_4(0,\sqrt{r'}))\),  
\begin{align}\label{eq:almost-tranf}
  |y|(\Delta\varphi)_K = |y|\big((W\varphi)_K-F_K\big)\,,
\end{align}
with
\begin{align}\label{eq:W-F}
  W(x)=\frac{W^{(1)}(x)}{|x|}+W^{(2)}(x)\,, \qquad
  F(x)=\frac{F^{(1)}(x)}{|x|}+F^{(2)}(x)\,.
\end{align}

Let \(f\in C_0^{\infty}(K^{-1}(\Omega))\); then there exists \(r'<r\)
such that \(\supp(f)\subset K^{-1}(\Omega')\), \(\Omega':=B_3(0,r')\);
choose  
 \(\{\varphi_n\}_{n\in\N}\subset C^{\infty}(\Omega')\) such that
\(\varphi_n\to\varphi\) and \(\Delta\varphi_n\to\Delta\varphi\) in 
\(L^2(\Omega')\)-norm. This is possible since $\varphi \in
W^{2,2}(\Omega')$. 
 Note that both \(\Delta 
 f\) and  \(4|y|^2f\) belong to \(C_0^{\infty}(K^{-1}(\Omega))\) when
 \(f\) does. 
Using \eqref{eq:Laplace-formula} for \(\varphi_n\in
 C^{\infty}(\Omega')\), 
Remark~\ref{rem:ext-K} in Section~\ref{sec:KS}
 below therefore implies that
\begin{align*}
  &\int_{K^{-1}(\Omega)} (\Delta f)(y)\,\varphi_{K}(y)\,dy
  =\lim_{n\to\infty} \int_{K^{-1}(\Omega)}(\Delta f)(y)
  \,(\varphi_{n})_{K}(y)\,dy
  \\&=\lim_{n\to\infty} \int_{K^{-1}(\Omega)}\!\!\!\!\!\!\!\!\!\!
  f(y)
  \,[\Delta_y(\varphi_{n})_{K}](y)\,dy
  =\lim_{n\to\infty} \int_{K^{-1}(\Omega)}\!\!\!\!\!\!\!\!\!\!
  4|y|^2f(y)
  \,(\Delta_x\varphi_{n})_{K}(y)\,dy
  \\&=\int_{K^{-1}(\Omega)}\!\!\!\!\! 
  4|y|^2f(y)
  \,(\Delta_x\varphi)_{K}(y)\,dy \,.
\end{align*}
It follows from this and \eqref{eq:almost-tranf} that
\begin{align*}
  &\int_{K^{-1}(\Omega)} (\Delta f)(y)\,\varphi_{K}(y)\,dy
  =\int_{K^{-1}(\Omega)} 4|y|^2f(y)\,\big((W\varphi)_K-F_K\big)(y)\,dy\,.
\end{align*}
Since \((W\varphi)_K=W_K\varphi_K\), and, by \eqref{eq:W-F} and \eqref{eq:norms}, 
\begin{align*}
  W_K(y)&=|y|^{-2}\big(W^{(1)}_K(y)+|y|^2W^{(2)}_K(y)\big)\,,\\
  F_K(y)&=|y|^{-2}\big(F^{(1)}_K(y)+|y|^2F^{(2)}_K(y)\big)\,,
\end{align*}
this implies that, for all \(f\in C_{0}^{\infty}(K^{-1}(\Omega))\), 
\begin{align*}
  \int_{K^{-1}(\Omega)}
  \varphi_K(y)\big[{}-&\Delta_yf(y)
  +4(W^{(1)}_K(y)+|y|^2W^{(2)}_K(y))f(y)\big]\,dy 
  \\&=\int_{K^{-1}(\Omega)} 4(F^{(1)}_K(y)+|y|^2F^{(2)}_K(y))f(y)\,dy\,,
\end{align*}
which means that \(\varphi_K\) satisfies \eqref{eq:new-psi-y} in the
distributional sense,  in \(K^{-1}(\Omega)=B_4(0,\sqrt{r})\).
\qed


\subsection{The \(N\)-particle problem}\label{sec:N}
In this section we prove Theorem~\ref{thm:main}. We only prove
the statement \eqref{eq:resultPSIa}, the proof of 
\eqref{eq:resultPSIb} is completely analogous, after an orthogonal
transformation of coordinates. We assume \(k=1\), the proof for other
\(k\)'s is the same.

Let \(H\) be given by \eqref{H}.
Then, with \((x,x')\equiv(x_1,x')\in\R^3\times\R^{3N-3}\),
\(x'=(x_2,\ldots,x_N)\),  
\begin{align}
  \label{eq:Hbis}
  H-E=-\Delta_{x}-\Delta_{x'}-\frac{Z}{|x|}+V_E(x,x')\,,
\end{align}
where 
\begin{align}\label{def:V-E}
  V_E(x_1,x')
  =\sum_{j=2}^N -\frac{Z}{|x_j|}+\sum_{1\le i<j\le
    N}\frac{1}{|x_i-x_j|} - E
\end{align}
is real analytic on \(\Omega\setminus\Sigma_1'\) (see
\eqref{eq:Sigma-k} for \(\Sigma_1'\)).

Assume \(\psi\in W^{1,2}(\Omega)\) satisfies 
\begin{align}\label{eq:new Schr}
  (H-E)\psi=0\quad \text{on} \quad\Omega\,,
\end{align}
%
and let \((x_0,x_0')\in\Omega\cap\Sigma_1\); then (see
\eqref{eq:sigmas}) \(x_0=0\).  We will first prove that there exists a
neighbourhood \(\Omega_1(P)\) of \(P=(0,x_0')\) and real analytic
functions \(\psi_{P}^{(1)},\psi_{P}^{(2)}:\Omega_1(P)\to\C\) such that 
\begin{align}
  \label{eq:resultPSIaBIS}
  &\psi({\bf x})=\psi_{P}^{(1)}({\bf x})+|x|\psi_{P}^{(2)}({\bf
    x})\,,\ {\bf x}\in\Omega_1(P)\,.
\end{align}

By the above, \(V_E\) is real analytic on a neighbourhood of \((0,x_0')\),
say, on 
\[
  U(R)=\big\{(x,x')\in\R^{3}\times\R^{3N-3}\,\big|\, |x|<R,
  |x'-x_0'|<R\big\}\subset\Omega
\] for some
\(R>0\), \(R\) small. Let
\begin{align}
  U_{K}(R):=\big\{(y,x')\in\R^{4}\times\R^{3N-3}\,\big|\, |y|<\sqrt{R},
  |x'-x_0'|<R\big\}\,. 
\end{align}
Define now, with \(K:\R^4\to\R^3\) as in \eqref{eq:KSbis},
\begin{align}
  \label{eq:def-new-u}
  u:U_{K}(R)\to\C\,&, \quad u(y,x'):=\psi(K(y),x')\,,\\
  \label{eq:def-new-w}
  W:U_{K}(R)\to\R\,&, \quad W(y,x'):=V_E(K(y),x')\,.
\end{align}
Since (by \eqref{eq:norms})
 \((K(y),x')\in U(R)\) for \((y,x')\in U_{K}(R)\),  it follows that \(u\)
 and \(W\) are well-defined,  and \(W\) is real analytic on \(U_{K}(R)\)
 since \(K\) is real analytic and \(V_E\) is real analytic on \(U(R)\).

As in the proof of Theorem~\ref{Hill}, we get that \eqref{eq:new Schr}
implies that \(u\) satisfies
\begin{align}
  \label{eq:new for u}
  Q(y,x',D_{y},D_{x'})u=0\quad \text{on} \quad U_{K}(R)\,,
\end{align}
where
\begin{align}\label{def:Grusin}
  Q(y,x',D_{y},D_{x'}):={}-\Delta_y-4|y|^2\Delta_{x'}+4|y|^2W(y,x')-4Z
\end{align}
is a degenerate elliptic operator, a
so-called `Gru{\v{s}}in-type operator'. 

Since \(|y|^2W(y,x')\) is real analytic on \(U_{K}(R)\), the operator
\(Q\) is (real) analytic hypoelliptic due to \cite[Theorem
5.1]{grusin2}. Therefore \eqref{eq:new for u} implies that \(u\) is
real analytic on some neighbourhood of \((0,x'_{0})\in\R^4\times\R^{3N-3}\).

It follows from Proposition~\ref{prop:many} in
Section~\ref{sec:AnAndKS} below that there exist a neighbourhood 
\(\Omega_1(P)\subset\R^{3N}\) of \(P=(0,x_0')\in\R^3\times\R^{3N-3}\)
and real analytic  
functions \(\psi_{P}^{(1)},\psi_{P}^{(2)}:\Omega_1(P)\to\C\) such that
\eqref{eq:resultPSIaBIS} holds.

Let now 
\begin{align*}
  \Omega_1:=\bigcup_{P\in\Omega\cap\Sigma_1}\Omega_{1}(P)\subset\Omega\subset\R^{3N}\,,
\end{align*}
and define \(\psi_{1}^{(1)}, \psi_{1}^{(2)}:\Omega_1\to\C\) by
\begin{align}\label{eq:def-psis}
  \psi_{1}^{(i)}({\bf x})=\psi_{P}^{(i)}({\bf x})\ \text{ when } {\bf x}
    \in\Omega_{1}(P)\ (i=1,2)\,.
\end{align}
To see that this is well-defined, we need to verify that if \({\bf
  x}\in \Omega_1(P)\cap\Omega_1(Q)\), then \(\psi_{P}^{(i)}({\bf
  x})=\psi_{Q}^{(i)}({\bf x})\), \(i=1,2\). Let therefore
\(\widetilde\psi^{(i)}=\psi^{(i)}_P-\psi^{(i)}_Q\), \(i=1,2\), then
\begin{align}\label{eq:unique-cont-assump}
  \widetilde\psi^{(1)}({\bf x})+|x|\widetilde\psi^{(2)}({\bf x})=0\,,\ 
  {\bf x}\in \Omega_1(P)\cap\Omega_1(Q)\,,
\end{align}
with  \(\widetilde\psi^{(1)}, \widetilde\psi^{(2)}\) real analytic in
\( \Omega_1(P)\cap\Omega_1(Q)\). Let \(\tilde{\bf x}_0=(0,\tilde{x}'_0)\in
\Omega_1(P)\cap\Omega_1(Q)\). Then, since \(\widetilde\psi^{(i)},
i=1,2\), are real analytic, there exist \(\delta>0\) and \(P_n^{(i)},
i=1,2\), homogeneous
polynomials of degree \(n\) such that 
\begin{align}\label{eq:expansion-here}
  \widetilde\psi^{(i)}({\bf x}) =
  \sum_{n=0}^{\infty}P_n^{(i)}(x,x'-\tilde{x}_0')\,, \quad i=1,2\,,
\end{align}
for \({\bf
  x}\in B_{3N}(\tilde{\bf x}_0,\delta)\). It follows from  
\eqref{eq:unique-cont-assump} (by homogeneity) that, for all
\(n\in\N\) and \({\bf x}=(x,x')\in B_{3N}(\tilde{\bf x}_0,\delta)\), 
\begin{align*}
  P^{(1)}_n(x, x'-\tilde{x}'_0)+|x| P^{(2)}_{n-1}(x,x'-\tilde{x}_0')=0\,. 
\end{align*}
But for \(n\) even, \(P^{(1)}_n\) is an even function,
while \(P_{n-1}^{(2)}\), and therefore \(|x|P^{(2)}_{n-1}\), is
odd. Therefore, \(P_n^{(1)}=P^{(2)}_{n-1}=0\). Similarly for \(n\)
odd. 
It follows that \(\widetilde\psi^{(1)}=\widetilde\psi^{(2)}=0\) on
\(B_{3N}(\tilde{\bf x}_0,\delta)\), and therefore also on
\(\Omega_1(P)\cap\Omega_1(Q)\), since \(\widetilde\psi^{(1)},
\widetilde\psi^{(2)}\) are real analytic. This proves that
\(\psi_{1}^{(1)}\) and \(\psi_{1}^{(2)}\) in
\eqref{eq:def-psis} are well-defined. Since they are obviously real
analytic, this finishes the proof of Thereom~\ref{thm:main}.
\qed
\section{The Kustaanheimo-Stiefel transform}
\label{sec:KS}
The KS-transform turns out to be a very useful and natural tool for the
investigation of Schr\"odinger equations with Coulombic
interactions. In particular \eqref{eq:norms} and the following lemma
are important for our proofs.  Most of the facts stated here are
well-known (see e.\ g.\ \cite[Appendix A]{HelfferEtAl}).
\begin{lem}\label{lem:2}
  Let \(K:\R^4\to\R^3\) be defined as in \eqref{eq:KSbis}, let
  \(f:\R^3\to\C\) be any \(C^2\)-function, and define
  \(f_K:\R^4\to\C\) by \eqref{eq:def-compo}.
  Finally, let 
  \begin{align}
   \label{def:P}
   L(y,D_y):=y_1\frac{\p}{\p y_4}-y_4\frac{\p}{\p y_1}
   +y_3\frac{\p}{\p y_2}-y_2\frac{\p}{\p y_3}\,.
   \end{align}
 
  \noindent{\rm (a)} Then, with \([A;B]=AB-BA\) the commutator of
  \(A\) and \(B\), 
   \begin{align}
    \label{eq:firstOrder}
    L(y,D_y)f_{K}&=0\ , \qquad\qquad \big[\Delta;L(y,D_y)\big]=0\,,
  \end{align}
  and \eqref{eq:Laplace-formula} holds.

\noindent{\rm (b)} Furthermore, for a function $g \in C^1({\mathbb
  R}^4)$, the following 
two statements are equivalent: 
\begin{enumerate}[{\rm (i)}]
\item \label{i} There exists a function $f:{\mathbb R}^3
  \rightarrow {\mathbb C}$
  such that $g= f_K$.
\item \label{ii} The function $g$ satisfies
  \begin{align}
    \label{eq:1}
    Lg = 0\,.
  \end{align}
\end{enumerate}

\noindent{\rm (c)} Finally, let \(U=B_3(0,r)\subset\R^3\) for
\(r\in(0,\infty]\). Then, for \(\phi\in C_0(\R^3)\) (continuous with
compact support),
\begin{align}
  \label{eq:L-2-formula}
  \int_{K^{-1}(U)}|\phi(K(y))|^2\,dy =
  \frac{\pi}{4}\int_{U}\frac{|\phi(x)|^2}{|x|}\,dx\,. 
\end{align}
In particular, 
\begin{align}
  \label{eq:L-2-abs}
  \big\||y|\phi_K\big\|^2_{L^2(K^{-1}(U))}=\frac{\pi}{4}\|\phi\|^2_{L^2(U)}\,.
\end{align}
\end{lem}
\begin{remark}\label{rem:ext-K}
By a density argument, the isometry \eqref{eq:L-2-abs} allows  to
extend the composition by \(K\) given by \eqref{eq:def-compo} (the
pull-back \(K^*\) by \(K\)) to a map
\begin{align*}
  K^{*}: L^2(U, dx)&\to L^2(K^{-1}(U), \tfrac{4}{\pi}|y|^2dy)\\
  \phi&\mapsto \phi_K
\end{align*}
in the case when \(U=B_3(0,r), r\in(0,\infty]\). 
This makes \(\phi_K\) well-defined for any \(\phi\in L^2(U)\). 
Furthermore, if \(\phi_n\to\phi\) in
\(L^2(U)\), then, for all \(g\in C^{\infty}(K^{-1}(U))\) (\(g\in
C_{0}^{\infty}(K^{-1}(U))\), if \(r=\infty\))
\begin{align}
  \label{eq:test-conv}
  \lim_{n\to\infty}\int_{K^{-1}(U)} g(y)(\phi_n)_K(y)\,dy
  =\int_{K^{-1}(U)} g(y)\phi_K(y)\,dy\,.
\end{align}
This follows from Schwarz' inequality and \eqref{eq:L-2-abs},
\begin{align*}
  \Big|\int_{K^{-1}(U)}&g(y)\big((\phi_n)_K(y)-\phi_K(y)\big)\,dy\Big|
  \\&\le
   \Big(\int_{K^{-1}(U)}\frac{|g(y)|^2}{|y|^2}\,dy\Big)^{1/2}
   \big\||y|\big((\phi_n)_K-\phi_K\big)\big\|_{L^2(K^{-1}(U))}
  \\&= \frac{\sqrt{\pi}}{2}\Big(\int_{K^{-1}(U)}\frac{|g(y)|^2}{|y|^2}\,dy\Big)^{1/2}
  \|\phi_n-\phi\|_{L^2(U)}\to0\,,\ n\to\infty\,.
\end{align*}
Here the \(y\)-integral clearly converges since \(g\in
C^{\infty}(\R^4)\) (\(g\in
C_{0}^{\infty}(\R^4)\), if \(r=\infty\)).

\end{remark}
\begin{remark}
  As a consequence of \eqref{eq:norms} and \eqref{eq:firstOrder}
  (choose 
  \(f(x)=|x|^j\)), we have that
  \begin{align}\label{eq:P norm zero} 
     L(y,D_y)|y|^{2j}=0\ ,\  j\in\N\,.
  \end{align}
\end{remark}

\begin{proof}[Proof of Lemma~\ref{lem:2}]
The lemma is easier to prove in `double polar coordinates' in
\(\R^4\). Let
\begin{align}
  \label{eq:doublepolar}
  (R,\Omega):=(r_1,r_2,\theta_1,\theta_2)
  \in(0,\infty)^2\times[0,2\pi)^2
\end{align}
be defined by the relation
\begin{align}
  \label{eq:polar coord}
     & y\equiv y(R,\Omega)=\big(y_1(R,\Omega), y_2(R,\Omega), y_3(R,\Omega),
     y_4(R,\Omega)\big)\,,\\
   &(y_1,y_4)=r_1(\cos\theta_1,\sin\theta_1)\ ,\quad
   (y_3,y_2)=r_2(\cos\theta_2,\sin\theta_2)\,.
   \label{eq:polar coordBIS}
\end{align}
Then it follows directly from \eqref{eq:KSbis} that 
\begin{align}
  \label{eq:K polar}
   K(y(R,\Omega))=\left(\begin{matrix}
   r_1^2-r_2^2\\
   -2r_1r_2\sin(\theta_1-\theta_2)\\
   2r_1r_2\cos(\theta_1-\theta_2)
   \end{matrix}\right)\,.
\end{align}

We note in passing that the relation \eqref{eq:norms} is immediate
from \eqref{eq:K polar}.
In the double polar coordinates, 
\begin{align}\label{eq:L in polar}
  L = \frac{\partial}{\partial \theta_1} + \frac{\partial}{\partial \theta_2}\,,
\end{align}
and 
\begin{align*}
  \Delta = \Big(\frac{\partial^2}{\partial r_1^2 } + \frac{1}{r_1}
  \frac{\partial}{\partial r_1 }+
\frac{1}{r_1^2} \frac{\partial^2}{\partial \theta_1^2}\Big) + 
\Big(\frac{\partial^2}{\partial r_2^2 } + \frac{1}{r_2}
  \frac{\partial}{\partial r_2 }+
\frac{1}{r_2^2} \frac{\partial^2}{\partial \theta_2^2}\Big)\,.
\end{align*}
Therefore, it is obvious that $L$ and $\Delta$ commute. Furthermore,
from \eqref{eq:K polar} we see that $f_K$ only depends on the angles
through the expression $\theta_1 - \theta_2$ and therefore,
\begin{align*}
  L f_K = \big( \frac{\partial}{\partial \theta_1} +
  \frac{\partial}{\partial \theta_2}\big) f_K = 0\,.
\end{align*}
The proof of \eqref{eq:Laplace-formula} is merely an elementary
computation, which we leave to the reader. This finishes the proof of
point (a) of the lemma.

From \eqref{eq:firstOrder} we infer that in order to prove  point (b)
we have to show that  \eqref{ii}
implies \eqref{i}.

To do so, we need to define a function \(f:\R^3\to\C\) such that
\(g=f_{K}\). If \(x=0\), let \(f(x):=g(0)\), then 
\(g(0)=f(0)=f(K(0))=f_{K}(0)\) by \eqref{eq:norms}. Assume now that
\(x\in\R^3\setminus\{0\}\). We claim that the pre-image of \(x\) under
\(K\), \(K^{-1}(\{x\})\), is a circle in \(\R^4\) (in the literature
called the `Hopf circle') 
and that \(g\) is
constant on this circle. Then, taking any \(y\in K^{-1}(\{x\})\) and
letting \(f(x):=g(y)\), we have that \(f\) is well-defined, and
satisfies \(f_{K}(y)=f(K(y))=f(x)=g(y)\).  This will finish the proof
of point (b) of the lemma.

To prove the claim, assume first that \(x\in\R^3\setminus\{0\}\),
\(x=(x_1,x_2,x_3)\) with \((x_2,x_3)\neq(0,0)\). Then the equations
(see \eqref{eq:K polar} and \eqref{eq:norms})
\begin{align}
  \label{eq:solve-K-invers}
   \begin{split} 
    r_1^2-r_2^2&=x_1, \\
   -2r_1r_2\sin\vartheta&=x_2,\\
   2r_1r_2\cos\vartheta&=x_3,\\
   (r_1^2+r_2^2)^2&=x_1^2+x_2^2+x_3^2
   \end{split}
\end{align}
uniquely determine \(r_1,r_2\in(0,\infty)\), and determine
\(\vartheta\) modulo \(2\pi\); choose the solution
\(\vartheta\in[0,2\pi)\).  That 
is, the pre-image of \(x\) under 
\(K\) is the set of points in \(\R^4\) with double polar coordinates
\((r_1,r_2,\theta_1,\theta_2)\), where \((r_1,r_2)\) is the unique
solution to \eqref{eq:solve-K-invers}, and
\(\theta_1-\theta_2=\vartheta\) modulo \(2\pi\), with
\(\vartheta\in[0,2\pi)\). Defining new angles  
\(\theta=\theta_1+\theta_2, \vartheta=\theta_1-\theta_2\), this set is
the circle in \(\R^4\) with centre at the origin and radius
\((x_1^2+x_2^2+x_3^2)^{1/4}=\sqrt{r_1^2+r_2^2}\), parametrized by
\(\theta\in[0,2\pi)\). Since, by \eqref{eq:L in polar}, the function
\(g\) (strictly speaking, \(g\) composed with the map in
\eqref{eq:polar coord}) is independent of \(\theta=\theta_1+\theta_2\),
\(g\) is, as claimed, constant on this circle. 

On the other hand, assume \(x=(t,0,0), t\in\R\setminus\{0\}\). Then
the equations
\begin{align}
  \label{eq:solve-K-invers-bis}
    \begin{split}
    r_1^2-r_2^2&=t, \\
     (r_1^2+r_2^2)^2&=t^2
    \end{split}
\end{align}
have a unique solution \((r_1,r_2)\); in fact,
\((r_1,r_2)=(\sqrt{t},0)\) if \(t>0\) and 
\((r_1,r_2)=(0,\sqrt{-t})\) if \(t<0\). In both cases, the pre-image of \(x\) under 
\(K\) is a circle, namely (see also \eqref{eq:doublepolar})
\begin{align*}
  C_{+}&=\{(\sqrt{t}\cos\theta_1,0,0,\sqrt{t}\sin\theta_1)\}\in\R^4\,|\,\theta_1\in[0,2\pi)\} 
  \quad\quad\ \,(t>0)\,,\\
  C_{-}&=\{(0,\sqrt{-t}\sin\theta_2,\sqrt{-t}\cos\theta_2,0)\}\in\R^4\,|\,\theta_2\in[0,2\pi)\}
  \quad (t>0)\,.
\end{align*}
Since \(y_2=y_3=0\) for any \(y=(y_1,y_2,y_3,y_4)\in C_{+}\),
\eqref{def:P} and \eqref{eq:1} imply that \(\p g/\p\theta_1=0\), with
\(\theta_1\) the angle parametrizing \(C_{+}\), and so \(g\) is, as
claimed, constant on \(C_{+}\); similarly for \(C_{-}\).  This finishes
the proof of point (b) of the lemma.

We finish by proving point (c); this is merely a calculation which we
for simplicity also do in `double polar coordinates': 
Recall that \(|y|^2=r_1^2+r_2^2=|x|\) (see \eqref{eq:norms}). 
By \eqref{eq:polar coordBIS} and \eqref{eq:K polar}, and since
\(U=B_3(0,r)\) and \(K^{-1}(U)=B_4(0,\sqrt{r})\),  
\begin{align*}
  \int_{K^{-1}(U)}\!\!\!\!\!\!\!
  |\phi(K(y))|^2\,dy=&\int_0^{2\pi}\Big\{\int_{0}^{\sqrt{r}}
  r_1\,dr_1\int_0^{\sqrt{r-r_1^2}} r_2\,dr_2\int_0^{2\pi}d\theta_1 \\
  \big|\phi\big(r_1^2-&r_2^2,-2r_1r_2\sin(\theta_1-\theta_2),
  2r_1r_2\cos(\theta_1-\theta_2)\big)\big|^2\Big\}\,d\theta_2\,.
\end{align*}
In the triple integral inside \(\big\{\cdot\}\) we make (for fixed
\(\theta_2\)) the change of
variables
\begin{align*}
  x=K_{\theta_2}(r_1,r_2,\theta_1)=\big(r_1^2-&r_2^2,-2r_1r_2\sin(\theta_1-\theta_2),
  2r_1r_2\cos(\theta_1-\theta_2)\big)\,.
\end{align*}
From the foregoing (see after \eqref{eq:solve-K-invers-bis}) it follows
that the image of \(K_{\theta_2}\) is \(U\). 
The determinant of the Jacobian is
\begin{align*}
  &\det(D K_{\theta_2})\\
  &=\left|\left[
   \begin{array}{ccc}
     2r_1 & -2r_2 & 0 \\
     -2r_2\sin(\theta_1-\theta_2) & -2r_1\sin(\theta_1-\theta_2) 
     & -2r_1r_2\cos(\theta_1-\theta_2)  \\
    2r_2\cos(\theta_1-\theta_2)  & 2r_1\cos(\theta_1-\theta_2)  &
   -2r_1r_2\sin(\theta_1-\theta_2)
   \end{array}
   \right]\right| \\
   & = 8r_1r_2(r_1^2+r_2^2)\,.
\end{align*}
Recall that \(|y|^2=r_1^2+r_2^2=|x|\). Therefore the integral is
\begin{align*}
  \int_{K^{-1}(U)}|\phi(K(y)|^2\,dy
  =\int_0^{2\pi}\Big\{\int_{U}|\phi(x)|^2\,\frac{dx}{8|x|}\Big\}\,d\theta_2 
  =\frac{\pi}{4}\int_{U}\frac{|\phi(x)|^2}{|x|}\,dx\,.
\end{align*}
This proves \eqref{eq:L-2-formula}; applying it to \(\sqrt{|x|}\phi\)
gives \eqref{eq:L-2-abs}. This finishes the proof of point (c), and
therefore, of Lemma~\ref{lem:2}.
\end{proof}
\begin{lemma}\label{lem:sph-harm}
Let the differential operator \(L=L(y,D)\) be given by \eqref{def:P},
and let $P_{2k}$ be a harmonic, homogeneous polynomial of degree $2k$ in
$\R^4$ such that $L P_{2k} = 0$. 

Then there exists a harmonic
polynomial in $\R^3$, $Y_{k}$, homogeneous of degree $k$, such that 
\begin{align}
  P_{2k}(y) = Y_{k}(K(y))\ \text{ for all } y\in \R^4\,,
\end{align}
with \(K:\R^4\to\R^3\) from \eqref{eq:KSbis}.
\end{lemma}
\begin{proof}
Using that $L P_{2k} = 0$ we get from Lemma~\ref{lem:2} the existence
of a function $Y_{k}$ such that $P_{2k}(y) = Y_{k}(K(y))$. Since the
KS-transform is homogeneous of degree $2$, $Y_{k}$ is necessarily
homogeneous of degree $k$. Furthermore, by \eqref{eq:Laplace-formula},
$Y_{k}$ is harmonic. So we only have left to prove that $Y_{k}$ is a
polynomial. 

Let ${\mathcal L}_n$ be the (positive) Laplace-Beltrami operator on
the sphere $\mathbb{S}^{n-1}$. Then one can express the Laplace
operator in 
$\R^n$ as 
\begin{align}\label{eq:Beltrami}
  \Delta = \frac{\p^2}{\p r^2} + \frac{n-1}{r} \frac{\p}{\p r} -
  \frac{{\mathcal L}_n}{r^2}\,. 
\end{align}
Furthermore, $\sigma({\mathcal L}_n) = \{
\ell(\ell+n-1)\}_{\ell=0}^{\infty}$ and the eigenspace corresponding
to the eigenvalue $\ell(\ell+n-1)$ is exactly spanned by the
restrictions to ${\mathbb S}^{n-1}$ of the harmonic, homogeneous
polymials in $\R^n$ of degree $\ell$.

Using the fact that $\Delta Y_{k}=0$ and that $Y_{k}$ is
homogeneous of degree $k$ in $\R^3$ we find that
$Y_{k}\big|_{{\mathbb S}^2}$ is an eigenfunction of ${\mathcal L}_3$
with eigenvalue $k(k+2)$. Thus there exists a homogeneous, harmonic
polynomial $\widetilde Y_{k}$ of degree $k$ such that 
\begin{align*}
  \widetilde Y_{k}\big|_{\mathbb{S}^2} = Y_{k}\big|_{\mathbb{S}^2}\,.
\end{align*}
Since the functions have the same homogeneity, they are identical
everywhere. This finishes the proof of the lemma. 
\end{proof}


\section{Analyticity and the KS-transform}
\label{sec:AnAndKS}

In this section we study the regularity of functions given as a
composition with the Kustaanheimo-Stiefel transform. We start with the
one-particle case. 

\begin{prop}\label{prop:one particle}
Let $U \subset \R^3$ be open with $0 \in U$, and let $\varphi:U\to \C$
be a function. Let ${\mathcal{U}} = K^{-1}(U) \subset \R^4$, with
\(K:\R^4\to\R^3\) from \eqref{eq:KSbis}, and suppose that
\begin{align}\label{assum:analy}
  \varphi_K = \varphi \circ K:\mathcal{U}\to \C
\end{align}
is real analytic. 

Then there exist functions $\varphi^{(1)}, \varphi^{(2)}$, real
analytic in a neighbourhood of $0 \in \R^3$, such that 
\begin{align}\label{res:prop}
  \varphi(x) = \varphi^{(1)}(x)+ |x|\varphi^{(2)}(x)\,.
\end{align}
\end{prop}
\begin{proof}
Note that \(K(-y)=K(y)\) for all \(y\in\R^4\), so that
\(\varphi_{K}(-y)=\varphi_{K}(y)\) for all \(y\in\R^4\). It follows
that \(\varphi_{K}\) can be written as an absolutely convergent power
series containing only terms of even order. Furthermore, since the sum
is absolutely convergent, the order of summation is unimportant, and
so, for some \(R>0\), \(c_{\beta}\in\C\),   
\begin{align}\label{eq:one}
   \varphi_{K}(y)=\sum_{\beta\in\N^4, |\beta|/2\in\N}c_\beta y^\beta
   =\sum_{n=0}^\infty\sum_{|\beta|=2n}c_\beta y^\beta
   \ \text{ for } \ |y|<R\,.
\end{align}
This implies (see e.\ g.\ \cite[sections 2.1--2.2]{Krantz}) that there
exists constants 
\(C_1,M_1>0\) such that
\begin{align}\label{eq:ja}
  |c_\beta|\le C_1 M_1^{|\beta|}\quad\text{for all } \beta\in\N^4\,. 
\end{align}

Note that for fixed \(n\in\N\), 
\begin{align}\label{def:Q}
  Q^{(2n)}(y):=\sum_{\beta\in\N^4,
  |\beta|=2n}c_{\beta}\,y^{\beta}
\end{align}
is a homogeneous polynomial of degree \(2n\). By \cite[Theorem
2.1]{Stein},
\begin{align}
  \label{eq:expansionP}
  Q^{(2n)}(y)=\sum_{j=0}^{n}|y|^{2j}H^{(2n)}_{2n-2j}(y)\,,
\end{align}
where \(H_{2n-2j}^{(2n)}\) is a homogeneous {\it harmonic} polynomial
of degree \(2n-2j\), \(j=0,1,\ldots,n\). It follows that
\begin{align}
  \label{eq:first sum}
  \varphi_{K}(y)=\sum_{n=0}^\infty Q^{(2n)}(y)
  =\sum_{n=0}^\infty\sum_{j=0}^{n}|y|^{2j}H^{(2n)}_{2n-2j}(y)\,.
\end{align}
We need the following lemma.
\begin{lemma}\label{lem:back to Y's}
There exist harmonic polynomials \(Y_{n-j}^{(2n)}:\R^3\to\C\),
homogeneous of degree \(n-j\), such that 
\begin{align}\label{eq: given by Y}
  H_{2n-2j}^{(2n)}(y) = Y_{n-j}^{(2n)}(K(y))\ \text{ for all } y\in \R^4\,,
\end{align}
with \(K:\R^4\to\R^3\) from \eqref{eq:KSbis}. In particular, the
function 
\begin{align}\label{eq:new q}
  q^{(2n)}(x):=\sum_{j=0}^{n}|x|^{j}Y_{n-j}^{(2n)}(x)
\end{align}
satisfies
\begin{align}\label{eq:relation}
  q^{(2n)}(K(y))=Q^{(2n)}(y) \ \text{ for all } y\in \R^4\,.
\end{align}
\end{lemma}
\begin{pf*}{Proof of Lemma~\ref{lem:back to Y's}}
Recall (see \eqref{eq:firstOrder}) that, with \(L\equiv L(y,D_y)\) as
in \eqref{def:P}, \(L\varphi_{K}=0\), and therefore, since power
series can be differentiated termwise (see \eqref{eq:first sum}), 
\begin{align}\label{eq:termbytermdiff}
  0=L\varphi_{K}=\sum_{n=0}^{\infty}LQ^{(2n)}\,.
\end{align}
Since \(LQ^{(2n)}\) is again a homogeneous polynomial of degree
\(2n\), it follows that 
\begin{align}\label{eq:PonQ}
  LQ^{(2n)}=0\,, \ n=0,1,\ldots\,.
\end{align}
Since \(L\) is a first order differential operator, \eqref{eq:P norm
  zero}  implies that 
\begin{align}\label{eq:compu P on terms}
  L[|y|^{2j}H_{2n-2j}^{(2n)}]
  =|y|^{2j}[LH_{2n-2j}^{(2n)}]\,,
\end{align}
where \(LH_{2n-2j}^{(2n)}\) is again a homogeneous polynomial of order
\(2(n-j)\). Then \eqref{eq:expansionP}, \eqref{eq:PonQ}, and
\eqref{eq:compu P on terms} imply that
\begin{align}
  \label{eq:zeroExp}
     \sum_{j=0}^{n}|y|^{2j}[LH_{2n-2j}^{(2n)}](y)=0\ \text{
       for all } y\in\R^4\,.
\end{align}
Since \(H_{2n-2j}^{(2n)}\) is harmonic, and (see \eqref{eq:firstOrder})
\([\Delta;L]=0\), we get that \(LH_{2n-2j}^{(2n)}\) is a homogeneous
{\it harmonic} polynomial of degree \(2(n-j)\). Note that for
\(|y|=1\), the left side of \eqref{eq:zeroExp} is a linear combination
of spherical harmonics of different degrees. From the linear
independence of such spherical harmonics it follows that 
\begin{align}\label{eq:PonPjzero}
  &LH_{2n-2j}^{(2n)}=0 \text{ for all } 
  j=0,\ldots,n, \text{ and }n\in\N\,.
\end{align}
From Lemma~\ref{lem:sph-harm} it follows that there exist harmonic
polynomials in $\R^3$, $Y_{n-j}^{(2n)}$, homogeneous of degree $n-j$,
such that \eqref{eq: given by Y} holds.

Now \eqref{eq:relation} follows from this and \eqref{eq:norms}. This
finishes the proof of the lemma. 
\end{pf*}
Lemma~\ref{lem:back to Y's}, \eqref{eq:first sum}, and
\(|K(y)|=|y|^2\), imply that 
\begin{align}
  \label{eq:first with K}
  \varphi_{K}(y)=\varphi(K(y))=\sum_{n=0}^{\infty}\sum_{j=0}^{n}
  |K(y)|^{j}\,Y_{n-j}^{(2n)}(K(y))\,. 
\end{align}
Formally, we can now finish the proof of Proposition~\ref{prop:one
  particle} by defining
\begin{align}\label{eq:phi1-temp}
  \varphi^{(1)}(x)&:=\sum_{n=0}^{\infty}\sum_{j=0,j\text{
      even}}^{n}\!\!|x|^{j}\,Y_{n-j}^{(2n)}(x)\,,\\
  \label{eq:phi2-temp}
  \varphi^{(2)}(x)&:=\sum_{n=0}^{\infty}\sum_{j=1,j\text{
      odd}}^{n}\!\!|x|^{j-1}\,Y_{n-j}^{(2n)}(x)\,.
\end{align}
However, it is not {\it a priori} clear that these sums converge and
thus define real analytic functions. The remainder of the proof will
establish the necessary convergence. 

\begin{lemma}\label{lem:split one particle}
There exists \(r>0\) such that the two series in \eqref{eq:phi1-temp}
and \eqref{eq:phi2-temp} converge for \(|x|<r\).  

More precisely, there exists a universal constant $R>0$ such that with
$\widetilde{C}_1:= R C_1$, $ \widetilde{M}_1 = 2 M_1^2$ (with $C_1,
M_1$ from \eqref{eq:ja}), 
\begin{align}\label{eq:BoundYn}
  |Y_{n-j}^{(2n)}(x) | \leq \widetilde{C}_1  \widetilde{M}_1^{n} |x|^{n-j}.
\end{align}
\end{lemma}
\begin{proof}
Clearly, the convergence of the series in \eqref{eq:phi1-temp} and
\eqref{eq:phi2-temp} is a consequence of \eqref{eq:BoundYn}: take $r <
1/(2 \widetilde{M}_1)$. Thus we only have to prove the estimate
\eqref{eq:BoundYn}. 

We return to \eqref{eq:one}. For fixed \(\beta\), with
\(|\beta|=2n>0\) we have (again using \cite[Theorem 2.1]{Stein}) that,
for some \(d_j^{(\beta)}\in\C\), 
\begin{align}\label{eq:two}
  y^\beta=\sum_{j=0}^{n}|y|^{2j}d_j^{(\beta)}P_{2n-2j}^{(\beta)}(y)\,,
\end{align}
where \(P_{2n-2j}^{(\beta)}\) is a harmonic homogeneous polynomial of
degree \(2n-2j\), which depends on \(\beta\), and satisfies 
\(\|P_{2n-2j}^{(\beta)}\|_{L^2(\mathbb{S}^3)}=1\).
It follows from \eqref{def:Q} and \eqref{eq:two} that
\begin{align}\label{eq:twoBis}
  Q^{(2n)}(y)=\sum_{j=0}^n |y|^{2j}\sum_{|\beta|=2n}c_\beta
  d_j^{(\beta)}P_{2n-2j}^{(\beta)}(y)\,.
\end{align}
Comparing \eqref{eq:expansionP} with \eqref{eq:twoBis} we see that
\begin{align}\label{new sums}
  \sum_{j=0}^{n}|y|^{2j}\big[H^{(2n)}_{2n-2j}(y)
  -\sum_{|\beta|=2n}c_{\beta}d_{j}^{(\beta)}P_{2n-2j}^{(\beta)}(y)\big]=0\,.
\end{align}
Restricting to \(|y|=1\), \eqref{new sums} becomes a sum of spherical
harmonics with different degrees, which are linearly independent,
implying that (see \eqref{eq: given by Y}) 
\begin{align}\label{eq:another new expansion}
  Y_{n-j}^{(2n)}(K(y))=H_{2n-2j}^{(2n)}(y)
  =\sum_{|\beta|=2n}c_{\beta}d_{j}^{(\beta)}P_{2n-2j}^{(\beta)}(y)\,. 
\end{align}

We are now going to bound the \(Y_{n-j}^{(2n)}\)'s in \(L^{\infty}\).
Since the (restriction to \(\mathbb{S}^3\) of the)
\(P_{2n-2j}^{(\beta)}\)'s in \eqref{eq:two} are orthogonal in
\(L^2(\mathbb{S}^3)\) (they are homogeneous of different degrees), we
get (by Parseval's identity), from setting \(|y|=1\) in
\eqref{eq:two}, that 
\begin{align}\label{eq:three}
  \sum_{j=0}^{n}|d_j^{(\beta)}|^2
  =\int_{\mathbb{S}^3}|y^\beta|^2\,d\omega  
  \le\int_{\mathbb{S}^3}|y|^{2|\beta|}\,d\omega 
  =\int_{\mathbb{S}^3}1\,d\omega=\Vol(\mathbb{S}^3)\,,
\end{align}
and so the \(d_j^{(\beta)}\)'s are bounded, uniformly in \(j\) and
\(\beta\), by \(\Vol(\mathbb{S}^3)^{1/2}\). 

Due to homogeneity, and using \cite[Lemma 8]{Mueller}, we get, for any
\(y\in\R^3\setminus\{0\}\) and \(j\le n\),  that  
\begin{align}\label{eq:eight}\nonumber
  \big|P_{2n-2j}^{(\beta)}(y)\big|
  &=|y|^{2n-2j}\big|P_{2n-2j}^{(\beta)}(y/|y|)\big|
  \le |y|^{2n-2j}\big\|P_{2n-2j}^{(\beta)}\big\|_{L^{\infty}(\mathbb{S}^3)}
  \\&\le |y|^{2n-2j}\frac{2n-2j+1}{\Vol(\mathbb{S}^3)^{1/2}}
  \le |y|^{2n-2j}\frac{3n}{\Vol(\mathbb{S}^3)^{1/2}}\,.
\end{align}
Note that (see \cite[pp.\ 138--139]{Stein})
\begin{align}\label{eq:no.multiindex}
  \# \big\{\sigma\in\N^k\,\big|\, |\sigma|=\ell\big\} =
  \binom{k+\ell-1}{k-1}\,, 
\end{align}
and so
\begin{align}\label{eq:non}
  \# \big\{\beta\in\N^4\,\big|\, |\beta|=2n\big\} &=
  \frac{(4+2n-1)!}{(4-1)!(2n)!} 
  \\&=\frac16(2n+3)(2n+2)(2n+1)\le 10n^3\,.\nonumber 
\end{align}
It follows from \eqref{eq:another new expansion}, \eqref{eq:ja},
\eqref{eq:three} \eqref{eq:eight}, and \eqref{eq:non} that (with
\(C_1\) and \(M_1\) the constants in \eqref{eq:ja}) 
\begin{align}\label{eq:estimateAgain}
  \big|Y_{n-j}^{(2n)}(K(y))\big|&\le
  \sum_{|\beta|=2n}|c_\beta|\,|d_j^{(\beta)}|\,\big|
  P_{2n-2j}^{(\beta)}(y)\big|
  \\&\le 10C_1n^4|y|^{2n-2j}M_1^{2n}
  =10C_1n^4|K(y)|^{n-j}M_1^{2n}\,.\nonumber
\end{align}
The desired estimate \eqref{eq:BoundYn} clearly follows, using the
surjectivity of $K$, with $R := 10 \max_{n}  n^4 2^{-n}$. 
\end{proof}

Recall that each term $|x|^j Y^{(2n)}_{n-j}(x)$ in the definition
\eqref{eq:phi1-temp} of
$\varphi^{(1)}$ is a homogeneous {\it polynomial} (of degree $n$) in
$x$, and similarly for $\varphi^{(2)}$. Therefore, the series
\eqref{eq:phi1-temp} and \eqref{eq:phi2-temp} are convergent power
series. This implies that $\varphi^{(1)}, \varphi^{(2)}$ define
real analytic functions on $\{ |x|< r\}$ (see \cite[sections 2.1--2.2]{Krantz}).

Finally, using \eqref{eq:first with K}, \eqref{eq:phi1-temp} and
\eqref{eq:phi2-temp},  
\begin{align}
  \label{eq:conv-series-again}\nonumber
  \varphi^{(1)}(K(y)) + &|K(y)|\varphi^{(2)}(K(y))
  \\&=\sum_{n=0}^{\infty}\sum_{j=0}^{n}|K(y)|^j\,Y^{(2n)}_{n-j}(K(y))
  =\varphi(K(y))\,.
\end{align}
This, and the surjectivity of $K$, imply \eqref{res:prop} and
therefore finishes the proof of Proposition~\ref{prop:one particle}. 
\end{proof}

For the \(N\)-particle case, we have the following analogous result. 
\begin{prop}\label{prop:many}
Let $U\subset \R^{3}$,  \(U'\subset\R^{3N-3}\) be open, with $0\in U$,
\(x_0'\in U'\) and let $\psi:U\times U' \to \C$ be a function. Let $\mathcal{U}
= K^{-1}(U) \subset \R^4$, with \(K:\R^4\to\R^3\) from
\eqref{eq:KSbis}, and suppose that   
\begin{align}\label{assum:analyN}
  u:\mathcal{U}\times U'&\to \C\\
   (y,x')&\mapsto \psi(K(y),x')\nonumber
\end{align}
is real analytic. 

Then there exist functions $\psi^{(1)}, \psi^{(2)}$, real analytic in
a neighbourhood \(\mathcal{W}\) of $(0,x_0') \in \R^{3N}$, such that  
\begin{align}\label{res:propN}
  \psi(x,x') = \psi^{(1)}(x,x')+ |x|\psi^{(2)}(x,x')\ ,
  \quad(x,x')\in \mathcal{W}\,.
\end{align}
\end{prop}
\begin{proof}
Define
\begin{align}\label{eq:IndexK}
  \varphi_{\gamma}(x) := \frac{1}{\gamma!} \partial_{x'}^{\gamma}
  \psi(x,x')\big|_{x'=x_0'}\, , \quad
  \varphi_{\gamma,K}(y):= \varphi_{\gamma}(K(y))\,.
\end{align}
This is well defined by the assumption on \(u\).

Since, as in the proof of Proposition~\ref{prop:one particle}, \(u\)
is even with respect to \(y\in\R^4\), and the series converges
absolutely, we have, for \(|y|<\sqrt{R}, |x'-x_0'|<R\) for some
\(R>0\), \(c_{\beta\gamma}\in\C\),
\begin{align*}
  u(y,x')&=\!\!\!\!
  \sum_{\beta\in\N^4,|\beta|/2\in\N,\gamma\in\N^{3N-3}}
  \!\!\!\!\!\!\!\!\!\!\!\!\!\!\!\!
  c_{\beta\gamma}\,y^{\beta}(x'-x_0')^{\gamma}\,,
\end{align*}
with
\begin{align}
  \label{eq:bounds-const-two}
  |c_{\beta\gamma}|\le
  C_2M_2^{|\beta+\gamma|}=C_2M_2^{|\beta|}M_2^{|\gamma|} 
  \quad\text{for all } \beta\in\N^4, \gamma\in\N^{3N-3}\,, 
\end{align}
for some constants \(C_2,M_2>0\). 
Clearly it follows that
\begin{equation}\label{eq:another-new}
  \varphi_{\gamma,K}(y) = \!\!\!\!\sum_{\beta\in\N^{4},|\beta|/2\in\N}
  \!\!\!\!c_{\beta\gamma}\,y^{\beta}\,,
\end{equation}
so that
\begin{align}
  \label{eq:expansion for new u}\nonumber
  u(y,x')&=\!\!\!\!\sum_{\gamma\in\N^{3N-3}}\!\!\!\big(\!\!\sum_{\beta\in\N^{4},|\beta|/2\in\N}  
  \!\!\!\!\!\!
  c_{\beta\gamma}\,y^{\beta}\big)(x'-x_0')^{\gamma}
  \\&=\sum_{\gamma\in\N^{3N-3}}\!\!\varphi_{\gamma,K}(y)\,(x'-x_0')^{\gamma}\,.
\end{align}
Moreover, from 
\eqref{eq:bounds-const-two} we have that, for all
\(\gamma\in\N^{3N-3}\), 
\begin{align}
  \label{eq:bounds-const-just-y}
  |c_{\beta\gamma}|\le C_1(\gamma)M_2^{|\beta|}\quad \text{ where }
  \quad C_1(\gamma):=C_2M_2^{|\gamma|}\,. 
\end{align}
In particular, \eqref{eq:another-new} and 
\eqref{eq:bounds-const-just-y} show that
$\varphi_{\gamma,K}$ is real analytic near $y=0$. Repeating
the arguments in the proof of  Proposition~\ref{prop:one particle} for
\(\varphi_{\gamma,K}\) for fixed \(\gamma\in\N^{3N-3}\), we get that 
\begin{align}\label{eq:some expansion}
  \varphi_{\gamma}(x)&= 
  \sum_{n=0}^{\infty}
  \sum_{\ell=0}^{[n/2]}|x|^{2\ell}\,Y^{(2n),\gamma}_{n-2\ell}(x)
  \\&\qquad+|x|\sum_{n=0}^{\infty}
  \sum_{\ell=0}^{[(n-1)/2]}|x|^{2\ell}\,
  Y^{(2n),\gamma}_{n-(2\ell+1)}(x)\,, \nonumber
\end{align}
where \(Y^{(2n),\gamma}_{n-k}:\R^3\to\C\) are harmonic polynomials,
homogeneous of degree \(n-k\), depending on \(\gamma\in\N^{3N-3}\). 
Therefore, for some \(a_{\alpha}(\gamma),b_{\alpha}(\gamma)\in\C\),
\(\alpha\in\N^{3}\),  
\begin{align}\label{eq:aalphas}
  \sum_{\ell=0}^{[n/2]}|x|^{2\ell}\,Y^{(2n),\gamma}_{n-2\ell}(x)
  &=\sum_{|\alpha|=n}a_{\alpha}(\gamma)x^{\alpha}\,,\\
  \label{eq:balphas}
  \sum_{\ell=0}^{[(n-1)/2]}|x|^{2\ell}\,Y^{(2n),\gamma}_{n-(2\ell+1)}(x)
  &=\sum_{|\alpha|=n-1}b_{\alpha}(\gamma)x^{\alpha}\,,
\end{align}
with (see \eqref{eq:BoundYn}), 
\begin{align}\label{eq:est-Y's-many-part}
  \big|\sum_{|\alpha|=n}a_{\alpha}(\gamma)x^{\alpha}\big|
  \leq R C_1(\gamma) n (2M_2^2)^n |x|^n,\\
  \big|\sum_{|\alpha|=n-1}b_{\alpha}(\gamma)x^{\alpha}\big|
  \le R C_1(\gamma)  n (2M_2^2)^n |x|^n.
\end{align}
Recall that (see \eqref{eq:no.multiindex})
\begin{align}\label{eq:no.mult.N}
   \# \{\gamma\in\N^{3N-3}\,|\, |\gamma|=k\} = \binom{3N+k-4}{3N-4}
   \,.
\end{align}
By definition, discarding part of the denominator,
\begin{equation*}
  \binom{3N+k-4}{3N-4}
  \leq
  \frac{(3N+k-4)!}{k!} =
  (3N+k-4)\cdot\ldots\cdot (k+1)\,.
\end{equation*}
This last product contains \((3N-4)\) terms each of which are smaller
than \((3N+k)\). Thus  
\begin{equation*}
  \binom{3N+k-4}{3N-4} \leq (3N+k)^{3N-4} \leq C_3 k^{3N}\,,
\end{equation*}
for some \(C_3\) (depending on $N$) and all \(k\geq 1\).

It follows that, for \(|x|<1/(4 M_2^2)\), \(|x'-x_0'|<1/(2M_2)\), 
\begin{align*}
  \Big|\sum_{\gamma\in\N^{3N-3}}&\sum_{n=0}^{\infty}
  \sum_{|\alpha|=n}a_{\alpha}(\gamma)x^{\alpha}(x'-x_0')^{\gamma}\Big|
  \\&\le R C_2
  \sum_{k=0}^{\infty}\sum_{\gamma\in\N^{3N-3},|\gamma|=k}\sum_{n=0}^{\infty}
  (2M_2^2)^n |x|^n
  M_2^{|\gamma|}|x'-x_0'|^{|\gamma|} 
  \\&\le R C_2C_3
  \Big(\sum_{k=0}^{\infty}\frac{k^{3N}}{2^{k}}\Big)
  \Big(\sum_{n=0}^{\infty}\frac{n}{2^n}\Big)<\infty\,,
\end{align*}
and so, with \(a_{\alpha\gamma}:=a_{\alpha}(\gamma)\), 
\begin{align}
  \label{eq:def-first-fct}
  \psi^{(1)}(x,x'):=
  \sum_{\gamma\in\N^{3N-3}}
  \sum_{\alpha\in\N^3}a_{\alpha\gamma}x^{\alpha}(x'-x_0')^{\gamma}
\end{align}
defines a real analytic function in a neighborhood of $(0,x_0')$.
Similarly, with \(b_{\alpha\gamma}:=b_{\alpha}(\gamma)\), 
\begin{align}
  \label{eq:def-second-fct}
  \psi^{(2)}(x,x'):=\sum_{\gamma\in\N^{3N-3}}
  \sum_{\alpha\in\N^3}b_{\alpha\gamma}x^{\alpha}(x'-x_0')^{\gamma}
\end{align}
defines a real analytic function in a neighborhood of $(0,x_0')$. From
the above observations and from \eqref{eq:expansion for new u},
\eqref{eq:IndexK}, \eqref{eq:some expansion}, \eqref{eq:aalphas} and
\eqref{eq:balphas} it follows that
\begin{align}
  \label{eq:second-to-last-form}\nonumber
  \psi^{(1)}&(K(y),x')+|K(y)|\,\psi^{(2)}(K(y),x')
  \\&=\nonumber
  \sum_{\gamma\in\N^{3N-3}}\sum_{n=0}^{\infty}
  \sum_{\ell=0}^{[n/2]}|K(y)|^{2\ell}\,
  Y^{(2n),\gamma}_{n-2\ell}(K(y))(x'-x_0')^{\gamma}
  \\&\ \ +\sum_{\gamma\in\N^{3N-3}}|K(y)|
  \nonumber
  \sum_{n=0}^{\infty}\sum_{\ell=0}^{[(n-1)/2]}|K(y)|^{2\ell}\,
  Y^{(2n),\gamma}_{n-(2\ell+1)}(K(y))(x'-x_0')^{\gamma}
  \\&=\sum_{\gamma\in\N^{3N-3}}\varphi_{\gamma,K}(y)(x'-x_0')^{\gamma} 
  =u(y,x')=\psi(K(y),x')\,,
\end{align}
and so, by the surjectivity of \(K\), 
\begin{align}
  \label{eq:last-formula}
  \psi(x,x')=\psi^{(1)}(x,x')+|x|\,\psi^{(2)}(x,x')\,,
\end{align}
with \(\psi^{(i)}, i=1,2\), real analytic on 
\begin{equation*} 
\big\{(x,x')\in\R^{3N}\,\big|\,|x|<1/(4 M_2^2)\,,\
|x'-x_0'|<1/(2M_2)\big\}\,. 
\end{equation*}

This finishes the proof of Proposition~\ref{prop:many}.
\end{proof}

\begin{acknowledgement} 
Financial support from the 
Danish Natural Science Research
Council, under the grant Mathematical Physics and Partial Differential
Equations (T\O S), and from
the European Science Foundation Programme {\it Spectral Theory and Partial 
  Differential Equations} (SPECT),
is gratefully acknowledged. SF is
supported by a Skou Grant and a Young Elite Researcher Award from the Danish Research
Council.
The authors wish to thank Bernard Helffer
(SF, T\O S),  G\"unther H\"ormann (MHO), Gerd Grubb (T\O S), Andreas
Knauf (T\O S), and Heinz Siedentop (T\O S)
for helpful discussions.
\end{acknowledgement} 

\providecommand{\bysame}{\leavevmode\hbox to3em{\hrulefill}\thinspace}
\providecommand{\MR}{\relax\ifhmode\unskip\space\fi MR }
\providecommand{\MRhref}[2]{%
  \href{http://www.ams.org/mathscinet-getitem?mr=#1}{#2}
}
\providecommand{\href}[2]{#2}

\end{document}